\documentclass[sigconf,nonacm]{acmart}

\AtBeginDocument{%
  }

\setcopyright{none}
\renewcommand\footnotetextcopyrightpermission[1]{}
\acmDOI{}
\acmISBN{}

\begin{document}

\title{UIGaze: How Closely Can VLMs Approximate Human Visual Attention on User Interfaces?}

\author{Min Song}
\email{elik@xebec.io}
\affiliation{%
  \institution{Xebec Inc.}
  \country{South Korea}}

\author{Yoonseong Lee}
\email{lordon@xebec.io}
\affiliation{%
  \institution{Xebec Inc.}
  \country{South Korea}}

\author{Yeonhu Seo}
\email{sean@xebec.io}
\affiliation{%
  \institution{Xebec Inc.}
  \country{South Korea}}

\renewcommand{\shortauthors}{Song et al.}

\begin{abstract}
Vision Language Models (VLMs) have demonstrated strong capabilities in understanding visual content, yet their ability to predict where humans look on user interfaces remains unexplored. We present UIGaze, a study investigating how closely VLMs can approximate human visual attention on user interfaces using real eye-tracking data. Using the UEyes dataset---comprising 1,980 UI screenshots across four categories (webpage, desktop, mobile, poster) with eye-tracking data from 62 participants---we evaluate nine state-of-the-art VLMs through a zero-shot coordinate prediction pipeline. Each model generates gaze point coordinates that are converted into saliency maps via Gaussian blurring and compared against ground truth using CC, SIM, and KL divergence. Our experiments (1,980 images $\times$ 9 models $\times$ 3 runs $\times$ 3 durations) reveal that VLMs achieve moderate alignment with human gaze patterns, with the degree of alignment varying significantly across UI types and improving with longer viewing durations---suggesting VLMs capture exploratory gaze patterns rather than initial fixations. All code, predictions, and evaluation results are publicly available.
\end{abstract}

\ccsdesc[500]{Human-centered computing~Empirical studies in HCI}
\ccsdesc[300]{Human-centered computing~User studies}

\keywords{Vision Language Models, Saliency Prediction, Eye Tracking, User Interface}

\maketitle

\section{Introduction}

Understanding where users direct their visual attention is fundamental to effective UI design. Saliency maps visualize these attention patterns and have become an essential tool in UX analysis, informing layout decisions, visual hierarchy, and the placement of critical interface elements. However, generating accurate saliency maps traditionally requires collecting real user data through eye-tracking studies or large-scale interaction analytics, both of which demand significant time, cost, and participant access. For products in pre-launch stages or with limited user bases, these approaches are effectively unusable.

Vision Language Models (VLMs) offer a potential alternative. With their ability to understand and reason about visual content, VLMs could predict where users are likely to look on a UI without requiring any real traffic. Meanwhile, dedicated saliency prediction models such as UMSI~\cite{fosco2020umsi} and DeepGaze~\cite{deepgaze2021} have shown strong performance on this task, but they require training on domain-specific eye-tracking data. VLMs, by contrast, can operate in a zero-shot setting---taking a UI screenshot and directly predicting gaze points without any prior training on saliency data.

In this paper, we present UIGaze, a study investigating to what extent VLMs can approximate human visual attention on user interfaces. We use the UEyes dataset~\cite{jiang2023ueyes}, which contains 1,980 UI screenshots across four categories (webpage, desktop, mobile, poster) with eye-tracking data from 62 participants, as our ground truth. Nine state-of-the-art VLMs are evaluated through a zero-shot coordinate prediction pipeline, where each model generates gaze point coordinates that are converted into saliency maps via Gaussian blurring and compared against ground truth using standard saliency metrics: CC, SIM, and KL divergence.

Our study reveals three key findings about how VLMs relate to human visual attention on user interfaces. We show that: (1) VLMs approximate exploratory attention but fundamentally fail to model initial fixations---predictions align significantly better with 7-second gaze patterns than 1-second; (2) the degree of behavioral alignment is shaped by UI structural characteristics, with clear visual hierarchies yielding stronger alignment than dense, competing layouts; and (3) UI interaction capability and attention prediction are fundamentally distinct---UI-TARS 1.5, designed for autonomous UI operation, shows near-zero correlation with human gaze. All code, model predictions, and evaluation results are publicly available at \url{https://github.com/dunward/uigaze}.

\section{Related Work}

\subsection{Saliency Prediction Models}

Computational saliency prediction has been driven primarily by deep learning approaches. UMSI~\cite{fosco2020umsi} proposed a unified model for predicting visual importance across graphic design types, while DeepGaze~\cite{deepgaze2021} leveraged transfer learning from ImageNet-trained backbones to achieve state-of-the-art fixation prediction. SAM~\cite{cornia2018sam} introduced an attentive mechanism for saliency map generation. These models achieve strong performance but require training on large-scale eye-tracking datasets. Jiang et al.~\cite{jiang2023ueyes} demonstrated that retraining UMSI on UI-specific eye-tracking data (UEyes) significantly improved its accuracy---from 0.778 to 0.878 AUC---highlighting the importance of domain-specific training for UI saliency prediction.

\subsection{UI Eye-Tracking Datasets}

Several datasets provide ground truth for visual saliency research. MIT300~\cite{mit300} and CAT2000~\cite{borji2015cat2000} are widely used benchmarks but focus on natural images rather than user interfaces. SALICON~\cite{jiang2015salicon} offers large-scale annotations on MS COCO images using mouse-contingent tracking as a proxy for eye movements, enabling cost-effective data collection at the expense of precision compared to dedicated eye trackers. UEyes~\cite{jiang2023ueyes} addresses the gap in UI-specific saliency data by providing eye-tracking recordings from 62 participants viewing 1,980 UI screenshots across four categories: webpage, desktop, mobile, and poster. It remains the most comprehensive publicly available eye-tracking dataset for user interfaces, with multi-duration saliency maps (1s, 3s, 7s) captured using a high-fidelity in-lab eye tracker.

\subsection{VLMs and Visual Saliency}

Recent work has begun to explore the intersection of VLMs and visual saliency. SalBench~\cite{salbench2025} evaluates whether VLMs can detect visually salient features such as odd-one-out patterns in synthetic images, finding that even GPT-4o achieves only 47.6\% accuracy on such tasks. While this task differs from saliency map prediction, it provides an early indication of VLMs' limitations in visual saliency understanding. SeeClick~\cite{screenspot2024}, which introduced the ScreenSpot benchmark, and similar approaches assess VLMs on UI element grounding---locating specific elements given natural language descriptions---which is a related but distinct task from predicting overall gaze distribution. However, no existing work investigates how closely VLMs can approximate human gaze distribution across an entire UI, using real eye-tracking data as ground truth. UIGaze addresses this gap by analyzing the behavioral alignment between VLM predictions and human attention patterns on the UEyes dataset.

\section{Methodology}

\subsection{Dataset}

We use the UEyes dataset~\cite{jiang2023ueyes}, a large-scale eye-tracking dataset comprising 1,980 UI screenshots across four categories: webpage, desktop UI, mobile UI, and poster, with 495 images per category. Eye-tracking data was collected from 62 participants using a high-fidelity in-lab eye tracker, with each participant viewing images for 7 seconds. The dataset provides ground truth saliency maps at three viewing durations---1 second, 3 seconds, and 7 seconds---representing progressively more exploratory gaze behavior. We evaluate against all three durations to examine how VLM predictions relate to different stages of visual attention.

\subsection{VLM Pipeline}

Our pipeline consists of three stages. First, a UI screenshot is sent to a VLM along with a zero-shot prompt instructing the model to predict 30--50 gaze points as a JSON array of normalized coordinates (x, y) with associated intensity values. Second, the predicted coordinates are projected onto a canvas matching the original image dimensions. Third, a Gaussian blur with $\sigma$=40px is applied to each point, weighted by its intensity, to produce a continuous saliency map. We adopt $\sigma$=40px to match the UEyes heatmap generation parameters, following the conventional approximation of 1\textdegree{} visual angle. We use a coordinate-based approach rather than image generation because generated images alter the original UI, making pixel-level comparison with ground truth infeasible.

\subsection{Models}

We evaluate nine VLMs from five providers, spanning both flagship and lightweight variants to examine how model capability relates to the degree of alignment with human attention. Table~\ref{tab:models} lists all models with their snapshot dates. All models are accessed through the OpenRouter API, providing a unified interface across providers.

\begin{table}[h]
\centering
\caption{VLMs evaluated in this study with their model snapshots.}
\label{tab:models}
\begin{tabular}{llc}
\toprule
\textbf{Provider} & \textbf{Model} & \textbf{Snapshot} \\
\midrule
OpenAI & GPT-5.4 & 2026-03-05 \\
OpenAI & GPT-5.4-mini & 2026-03-17 \\
Google & Gemini 3.1 Pro & 2026-02-19 \\
Google & Gemini 3.1 Flash Lite & 2026-03-03 \\
Anthropic & Claude Opus 4.6 & 2026-02-05 \\
Anthropic & Claude Sonnet 4.6 & 2026-02-17 \\
Qwen & Qwen 3.5 Plus & 2026-02-16 \\
Qwen & Qwen 3.5 Flash & 2026-02-24 \\
ByteDance & UI-TARS 1.5 & --$^*$ \\
\bottomrule
\end{tabular}

\vspace{0.5em}

\noindent\small{$^*$ No snapshot date available in the model identifier.}
\end{table}

\subsection{Evaluation Metrics}

We use three standard saliency evaluation metrics. CC (Correlation Coefficient) measures the linear correlation between predicted and ground truth saliency maps, where values closer to 1 indicate higher similarity in spatial pattern. SIM (Similarity) treats both maps as probability distributions and computes their overlap, with 1 indicating identical distributions. KL (KL Divergence) quantifies the information loss when the ground truth distribution is approximated by the predicted distribution, where values closer to 0 indicate stronger alignment.

\subsection{Experimental Setup}

Each of the 1,980 images is evaluated with all 9 models, with 3 independent runs per model-image pair to account for VLM output variability. We conducted a pilot study with 10 runs on a subset of 40 images, which showed negligible difference in mean metrics between 3 and 10 runs. Based on this, we use 3 runs for the full experiment. Each prediction is compared against ground truth saliency maps at all three durations (1s, 3s, 7s). We report mean and standard deviation across runs for each metric. Experiments were conducted in April 2026.

\section{Results}

\subsection{Overall Performance}

Table~\ref{tab:overall} presents how closely each model's predictions align with human gaze patterns at 7-second viewing duration. GPT-5.4 shows the strongest alignment with human gaze patterns across all three metrics (CC=0.408, SIM=0.503, KL=1.345), followed by Claude Opus 4.6 (CC=0.344) and Qwen 3.5 Plus (CC=0.337).\footnote{For Qwen 3.5 Plus and Qwen 3.5 Flash, two images were excluded due to consistent response failures; metrics for these models are computed on the remaining 1,978 images.} A clear gap separates the top-tier models from those showing minimal alignment with human attention: Gemini 3.1 Pro (CC=0.144), Gemini 3.1 Flash Lite (CC=0.105), and UI-TARS 1.5 (CC=0.086) all remain below CC=0.15, with KL divergence values exceeding 5.0, indicating predictions that substantially diverge from human gaze patterns.

\begin{table*}[h]
\centering
\caption{Overall saliency prediction performance at 7s duration (mean $\pm$ std). Best results in \textbf{bold}.}
\label{tab:overall}
\begin{tabular}{lccc}
\toprule
\textbf{Model} & \textbf{CC $\uparrow$} & \textbf{SIM $\uparrow$} & \textbf{KL $\downarrow$} \\
\midrule
\textbf{GPT-5.4} & \textbf{0.408} $\pm$ 0.169 & \textbf{0.503} $\pm$ 0.085 & \textbf{1.345} $\pm$ 0.793 \\
Claude Opus 4.6 & 0.344 $\pm$ 0.200 & 0.468 $\pm$ 0.102 & 1.670 $\pm$ 0.957 \\
Qwen 3.5 Plus & 0.337 $\pm$ 0.167 & 0.466 $\pm$ 0.085 & 1.612 $\pm$ 0.869 \\
Qwen 3.5 Flash & 0.295 $\pm$ 0.171 & 0.424 $\pm$ 0.102 & 2.214 $\pm$ 1.372 \\
GPT-5.4-mini & 0.289 $\pm$ 0.171 & 0.453 $\pm$ 0.084 & 1.600 $\pm$ 0.918 \\
Claude Sonnet 4.6 & 0.279 $\pm$ 0.195 & 0.436 $\pm$ 0.110 & 1.924 $\pm$ 1.247 \\
Gemini 3.1 Pro & 0.144 $\pm$ 0.228 & 0.255 $\pm$ 0.196 & 5.457 $\pm$ 3.385 \\
Gemini 3.1 Flash Lite & 0.105 $\pm$ 0.242 & 0.234 $\pm$ 0.216 & 5.823 $\pm$ 3.871 \\
UI-TARS 1.5 & 0.086 $\pm$ 0.171 & 0.142 $\pm$ 0.143 & 7.448 $\pm$ 2.580 \\
\bottomrule
\end{tabular}
\end{table*}

\subsection{Effect of Viewing Duration}

Table~\ref{tab:duration} compares behavioral alignment across the three viewing durations. All models exhibit stronger alignment with human attention as viewing duration increases. GPT-5.4 improves from CC=0.217 (1s) to 0.326 (3s) to 0.408 (7s), representing an 88\% relative increase. This pattern holds across all models, indicating that VLMs approximate the deliberate, semantically-driven scanning behavior that emerges over time rather than the reflexive orienting responses that dominate early viewing. Notably, model rankings shift across durations: Claude Opus 4.6 ranks first at 1s (CC=0.227) but is overtaken by GPT-5.4 at 3s and 7s, suggesting that different models may capture different temporal aspects of human visual attention.

\begin{table*}[h]
\centering
\caption{CC scores across viewing durations (mean $\pm$ std). Best results per duration in \textbf{bold}.}
\label{tab:duration}
\begin{tabular}{lccc}
\toprule
\textbf{Model} & \textbf{1s} & \textbf{3s} & \textbf{7s} \\
\midrule
GPT-5.4 & 0.217 $\pm$ 0.143 & \textbf{0.326} $\pm$ 0.163 & \textbf{0.408} $\pm$ 0.169 \\
Claude Opus 4.6 & \textbf{0.227} $\pm$ 0.157 & 0.303 $\pm$ 0.181 & 0.344 $\pm$ 0.200 \\
Qwen 3.5 Plus & 0.161 $\pm$ 0.131 & 0.245 $\pm$ 0.162 & 0.337 $\pm$ 0.167 \\
Qwen 3.5 Flash & 0.135 $\pm$ 0.137 & 0.210 $\pm$ 0.162 & 0.295 $\pm$ 0.171 \\
GPT-5.4-mini & 0.154 $\pm$ 0.133 & 0.220 $\pm$ 0.161 & 0.289 $\pm$ 0.171 \\
Claude Sonnet 4.6 & 0.142 $\pm$ 0.149 & 0.213 $\pm$ 0.178 & 0.279 $\pm$ 0.195 \\
Gemini 3.1 Pro & 0.059 $\pm$ 0.140 & 0.103 $\pm$ 0.189 & 0.144 $\pm$ 0.228 \\
Gemini 3.1 Flash Lite & 0.038 $\pm$ 0.147 & 0.070 $\pm$ 0.196 & 0.105 $\pm$ 0.242 \\
UI-TARS 1.5 & 0.078 $\pm$ 0.153 & 0.101 $\pm$ 0.180 & 0.086 $\pm$ 0.171 \\
\bottomrule
\end{tabular}
\end{table*}

\subsection{Performance by UI Type}

Table~\ref{tab:category} shows how behavioral alignment varies across UI categories at 7s duration. Human attention patterns on desktop UIs are most accurately approximated by VLMs, with GPT-5.4 reaching CC=0.506---the strongest single result across all model-category combinations. Alignment patterns vary across models: while GPT-5.4 performs best on desktop, the relative ordering of other categories differs by model. Web UIs are not the hardest category for top-tier models---for GPT-5.4, web (CC=0.371) is comparable to poster (CC=0.388) and mobile (CC=0.369). However, mobile UIs prove particularly challenging for certain models: UI-TARS 1.5 produces predictions inversely correlated with actual human gaze on mobile (CC=$-$0.008), suggesting a fundamental misalignment between its UI interaction training and human visual attention behavior. Gemini models show uniformly weak alignment across all categories, with KL values consistently above 4.0.

\begin{table*}[h]
\centering
\caption{CC scores by UI category at 7s duration (mean $\pm$ std). Best results per category in \textbf{bold}.}
\label{tab:category}
\begin{tabular}{lcccc}
\toprule
\textbf{Model} & \textbf{Desktop} & \textbf{Mobile} & \textbf{Poster} & \textbf{Web} \\
\midrule
\textbf{GPT-5.4} & \textbf{0.506} $\pm$ 0.156 & \textbf{0.369} $\pm$ 0.165 & \textbf{0.388} $\pm$ 0.169 & \textbf{0.371} $\pm$ 0.145 \\
Claude Opus 4.6 & 0.404 $\pm$ 0.193 & 0.275 $\pm$ 0.210 & 0.371 $\pm$ 0.183 & 0.325 $\pm$ 0.191 \\
Qwen 3.5 Plus & 0.402 $\pm$ 0.162 & 0.325 $\pm$ 0.158 & 0.323 $\pm$ 0.182 & 0.298 $\pm$ 0.146 \\
GPT-5.4-mini & 0.399 $\pm$ 0.168 & 0.230 $\pm$ 0.148 & 0.255 $\pm$ 0.177 & 0.272 $\pm$ 0.134 \\
Claude Sonnet 4.6 & 0.393 $\pm$ 0.180 & 0.187 $\pm$ 0.203 & 0.270 $\pm$ 0.167 & 0.265 $\pm$ 0.167 \\
Qwen 3.5 Flash & 0.359 $\pm$ 0.185 & 0.312 $\pm$ 0.163 & 0.246 $\pm$ 0.168 & 0.263 $\pm$ 0.145 \\
Gemini 3.1 Pro & 0.220 $\pm$ 0.258 & 0.134 $\pm$ 0.213 & 0.129 $\pm$ 0.208 & 0.093 $\pm$ 0.211 \\
Gemini 3.1 Flash Lite & 0.113 $\pm$ 0.258 & 0.112 $\pm$ 0.250 & 0.186 $\pm$ 0.240 & 0.007 $\pm$ 0.179 \\
UI-TARS 1.5 & 0.108 $\pm$ 0.171 & $-$0.008 $\pm$ 0.111 & 0.169 $\pm$ 0.173 & 0.076 $\pm$ 0.171 \\
\bottomrule
\end{tabular}
\end{table*}

\subsection{Qualitative Examples}

\begin{figure*}[h]
\centering
\includegraphics[width=\textwidth]{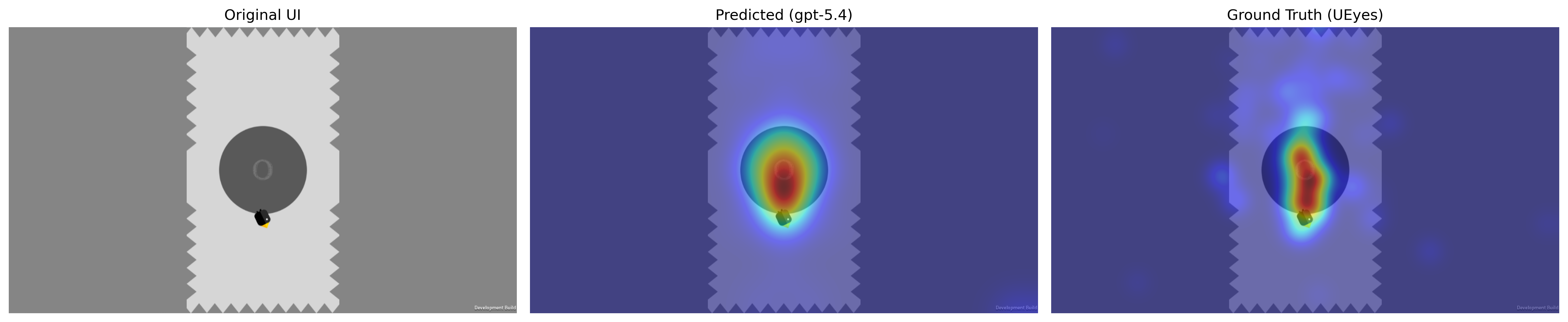}
\vspace{0.5cm}
\includegraphics[width=\textwidth]{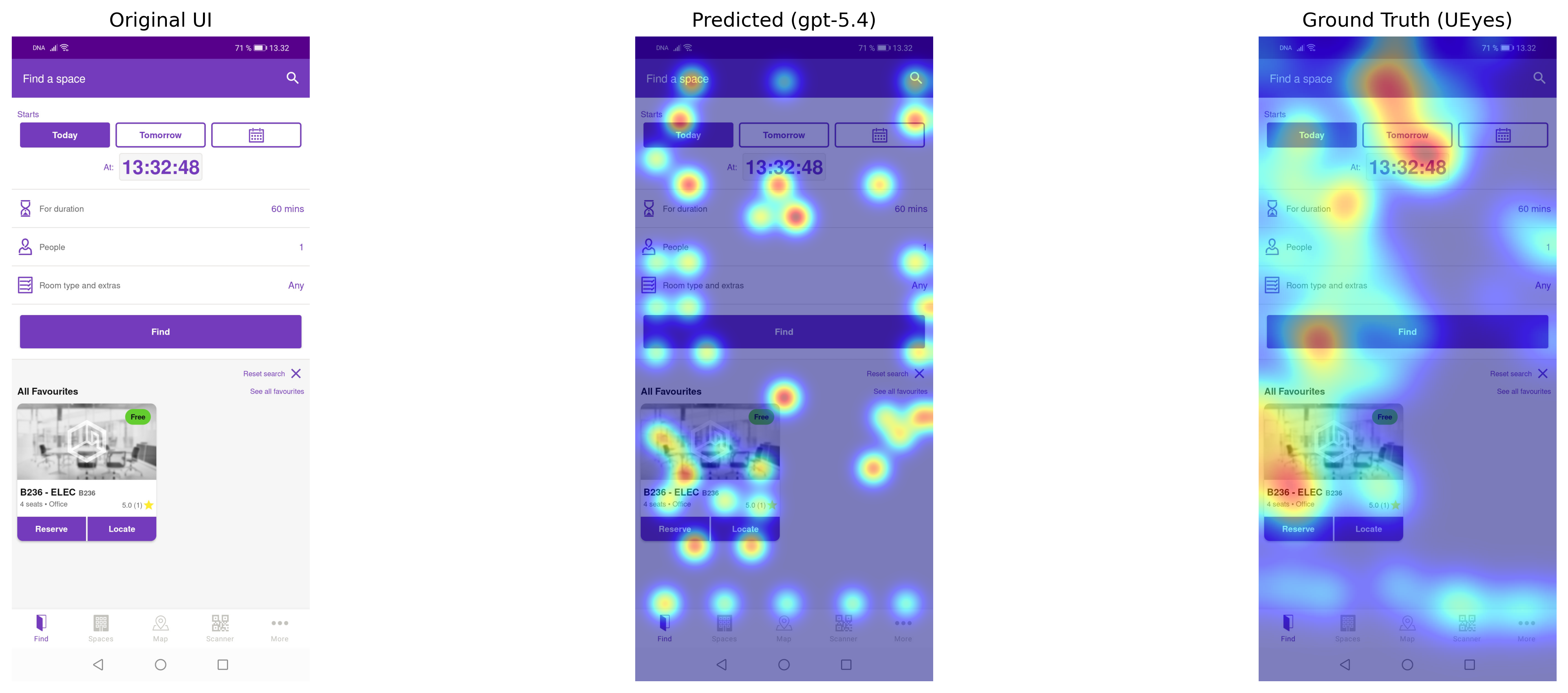}
\caption{Best (top) and worst (bottom) saliency prediction cases for GPT-5.4 at 7s duration. Each row shows the original UI (left), VLM-predicted saliency map (center), and ground truth from UEyes (right).}
\Description{Two rows of comparison images. Top row shows a desktop UI where the VLM prediction closely matches the ground truth heatmap. Bottom row shows a mobile UI where the VLM prediction diverges significantly from the ground truth.}
\label{fig:qualitative}
\end{figure*}

Figure~\ref{fig:qualitative} presents representative best and worst prediction cases from GPT-5.4 at 7s duration. In the best case (desktop UI), the VLM-generated saliency map closely matches human gaze patterns, with attention concentrated on primary UI elements such as headings, buttons, and navigation areas. This desktop interface features a clear visual hierarchy with well-separated functional regions, allowing the VLM to correctly identify the dominant areas of human attention. In the worst case (mobile UI), VLM predictions diverge significantly from actual human viewing behavior. The mobile interface presents a dense, vertically scrolling layout with multiple interactive elements competing for attention in a narrow viewport. The VLM distributes predicted gaze broadly across semantic elements, while the human ground truth reveals more focused fixation patterns on specific interactive components. This contrast illustrates a key limitation: VLMs reason about what is semantically important, but human gaze is also driven by visual salience, layout density, and interaction affordances that VLMs do not explicitly model.

\section{Discussion}

\subsection{Duration and VLM Prediction Alignment}

The consistent improvement across all models from 1s to 7s ground truth suggests that VLMs predict attention patterns more similar to exploratory viewing than to initial fixations. This is intuitive---VLMs analyze the full image and identify semantically important elements, which aligns more closely with where users look after they have had time to explore an interface rather than where their eyes land first. Interestingly, model rankings shift across durations: Claude Opus 4.6 leads at 1s but GPT-5.4 overtakes at 3s and 7s, suggesting that different models may capture different temporal aspects of visual attention. UI-TARS 1.5 is a notable exception---its alignment with human gaze barely changes across durations (CC: 0.078 $\rightarrow$ 0.101 $\rightarrow$ 0.086), indicating a fundamental inability to predict human gaze patterns regardless of viewing time.

\subsection{Variation Across UI Types}

GPT-5.4 achieves its strongest behavioral alignment on desktop UIs (CC=0.506), while its alignment on mobile (0.369), poster (0.388), and web (0.371) is comparatively similar. This does not support a simple ``simpler UIs are easier to predict'' narrative. Instead, our results suggest that alignment quality is influenced by specific UI structural characteristics. Desktop interfaces typically feature a clear visual hierarchy with distinct navigation bars, content areas, and sidebars, which may provide strong spatial cues that VLMs can leverage. Mobile interfaces, by contrast, present dense layouts with many competing elements in a narrow viewport, making gaze distribution harder to approximate. Poster designs, despite their visual simplicity, distribute attention across decorative elements in ways that may not align with how VLMs prioritize semantic content. Different models show different category strengths: Qwen 3.5 Flash achieves relatively strong alignment on mobile (CC=0.312) compared to its poster performance (CC=0.246), while Claude Opus shows the opposite pattern. UI-TARS 1.5 produces predictions inversely correlated with actual human gaze on mobile (CC=$-$0.008), suggesting a fundamental misalignment between its UI interaction training and human visual attention behavior.

\subsection{Limitations of UI-Specialized Models}

UI-TARS 1.5, despite being designed specifically for UI interaction tasks such as clicking, tapping, and navigating interfaces, ranks last among all evaluated models. Its KL divergence ranges from 5.4 to 8.8 across categories, far worse than general-purpose VLMs. This highlights an important distinction: performing actions on UI elements (e.g., clicking buttons or filling forms) is a fundamentally different capability from predicting where humans direct their visual attention. UI-TARS excels at autonomous UI operation---executing action sequences on interfaces given natural language instructions---but this task requires identifying actionable elements rather than modeling the holistic distribution of human visual attention.

\subsection{Zero-shot VLMs vs.\ Trained CNN Models}

Jiang et al.\ reported that UMSI, when retrained on UEyes data, achieved an AUC of 0.878---a strong result enabled by domain-specific training. In contrast, the VLM showing strongest alignment with human gaze in our study (GPT-5.4) achieves CC=0.408 at 7s duration in a fully zero-shot setting with no exposure to eye-tracking data. While these metrics are not directly comparable, the gap illustrates the cost of foregoing domain-specific training. Nevertheless, VLMs offer a practical advantage: they require no training data, no model fine-tuning, and can be applied immediately to any UI via a simple API call. For scenarios where approximate saliency estimates are sufficient---such as early-stage design review or rapid prototyping---VLMs may provide a useful, if imperfect, alternative to dedicated saliency models.

\subsection{Practical Implications}

Beyond academic analysis, our findings suggest practical applications for VLM-based saliency prediction. Even at moderate alignment levels, zero-shot VLM predictions could serve as a lightweight proxy for human attention during early-stage design processes. Designers could use VLM saliency maps to identify potential attention blind spots before committing to user testing, conduct rapid sanity checks on UI layouts prior to A/B testing, or approximate gaze behavior for interfaces that have not yet been deployed. While VLM predictions should not replace formal user studies, they offer a cost-effective preliminary signal that requires no participant recruitment, no specialized equipment, and no data collection infrastructure.

\section{Limitations and Future Work}

Our finding that VLMs approximate exploratory attention but not initial fixations opens a specific research question: can prompt engineering or few-shot examples narrow this gap? The current study uses a single zero-shot prompt, and prompt design choices such as the number of requested points or the inclusion of temporal cues (e.g., ``predict where users look within the first second'') may meaningfully shift predictions toward earlier attention stages.

The UEyes ground truth aggregates gaze data across all 62 participants, which captures population-level attention tendencies but obscures individual variation. This limitation is particularly relevant for UI design, where target user demographics may differ significantly. Collecting gaze data segmented by age, expertise, or cultural background would enable evaluation of whether VLMs can differentially approximate the attention patterns of specific user groups. A promising direction is persona-based prompting, where VLMs are instructed to simulate the gaze behavior of a particular user demographic without requiring new eye-tracking data collection.

Our evaluation uses three standard saliency metrics (CC, SIM, KL). Additional metrics such as AUC-Judd and NSS could reveal aspects of alignment not captured by distributional measures, particularly regarding whether VLMs correctly identify the most fixated regions even when overall spatial distributions differ.

Finally, a controlled comparison between zero-shot VLMs and trained CNN saliency models (e.g., UMSI++) using identical metrics on the same dataset would quantify the precise cost of foregoing domain-specific training, and clarify whether VLM alignment with human attention can be improved through fine-tuning or few-shot learning on eye-tracking data.

\section{Conclusion}

We presented UIGaze, a study investigating how closely nine VLMs can approximate human visual attention on user interfaces, using real eye-tracking data from the UEyes dataset. Our experiments reveal that VLMs achieve moderate alignment with human gaze patterns in a fully zero-shot setting, with GPT-5.4 reaching CC=0.408 at 7-second duration. Alignment improves consistently with longer viewing durations across all models, suggesting VLMs capture exploratory gaze behavior rather than initial fixations. We also find that alignment varies across UI types and that UI interaction-specialized models do not necessarily excel at approximating human attention. While a significant gap remains between zero-shot VLMs and trained saliency models, VLMs offer a practical advantage in requiring no training data and being immediately applicable to any interface. All code, model predictions, and evaluation results are publicly available at \url{https://github.com/dunward/uigaze} to support future research in this direction.

\bibliographystyle{ACM-Reference-Format}
\bibliography{references}


\begin{thebibliography}{9}


\ifx \showCODEN    \undefined \def \showCODEN     #1{\unskip}     \fi
\ifx \showISBNx    \undefined \def \showISBNx     #1{\unskip}     \fi
\ifx \showISBNxiii \undefined \def \showISBNxiii  #1{\unskip}     \fi
\ifx \showISSN     \undefined \def \showISSN      #1{\unskip}     \fi
\ifx \showLCCN     \undefined \def \showLCCN      #1{\unskip}     \fi
\ifx \shownote     \undefined \def \shownote      #1{#1}          \fi
\ifx \showarticletitle \undefined \def \showarticletitle #1{#1}   \fi
\ifx \showURL      \undefined \def \showURL       {\relax}        \fi
\providecommand\bibfield[2]{#2}
\providecommand\bibinfo[2]{#2}
\providecommand\natexlab[1]{#1}
\providecommand\showeprint[2][]{arXiv:#2}

\bibitem[Borji and Itti(2015)]%
        {borji2015cat2000}
\bibfield{author}{\bibinfo{person}{Ali Borji} {and} \bibinfo{person}{Laurent Itti}.} \bibinfo{year}{2015}\natexlab{}.
\newblock \showarticletitle{CAT2000: A Large Scale Fixation Dataset for Boosting Saliency Research}.
\newblock \bibinfo{journal}{\emph{arXiv preprint arXiv:1505.03581}} (\bibinfo{year}{2015}).
\newblock


\bibitem[Cheng et~al\mbox{.}(2024)]%
        {screenspot2024}
\bibfield{author}{\bibinfo{person}{Kanzhi Cheng}, \bibinfo{person}{Qiushi Sun}, \bibinfo{person}{Yougang Chu}, \bibinfo{person}{Fangzhi Xu}, \bibinfo{person}{Yantao Li}, \bibinfo{person}{Jianbing Zhang}, {and} \bibinfo{person}{Zhiyong Wu}.} \bibinfo{year}{2024}\natexlab{}.
\newblock \showarticletitle{SeeClick: Harnessing GUI Grounding for Advanced Visual GUI Agents}. In \bibinfo{booktitle}{\emph{arXiv preprint arXiv:2401.10935}}.
\newblock


\bibitem[Cornia et~al\mbox{.}(2018)]%
        {cornia2018sam}
\bibfield{author}{\bibinfo{person}{Marcella Cornia}, \bibinfo{person}{Lorenzo Baraldi}, \bibinfo{person}{Giuseppe Serra}, {and} \bibinfo{person}{Rita Cucchiara}.} \bibinfo{year}{2018}\natexlab{}.
\newblock \showarticletitle{Predicting Human Eye Fixations via an LSTM-based Saliency Attentive Model}.
\newblock \bibinfo{journal}{\emph{IEEE Transactions on Image Processing}} \bibinfo{volume}{27}, \bibinfo{number}{10} (\bibinfo{year}{2018}).
\newblock


\bibitem[Dahou et~al\mbox{.}(2025)]%
        {salbench2025}
\bibfield{author}{\bibinfo{person}{Yasser Dahou}, \bibinfo{person}{Ngoc~Dung Huynh}, \bibinfo{person}{Phuc~H. Le-Khac}, \bibinfo{person}{Wamiq~Reyaz Para}, \bibinfo{person}{Ankit Singh}, {and} \bibinfo{person}{Sanath Narayan}.} \bibinfo{year}{2025}\natexlab{}.
\newblock \showarticletitle{Vision-Language Models Can't See the Obvious}. In \bibinfo{booktitle}{\emph{Proceedings of the IEEE/CVF International Conference on Computer Vision (ICCV)}}.
\newblock


\bibitem[Fosco et~al\mbox{.}(2020)]%
        {fosco2020umsi}
\bibfield{author}{\bibinfo{person}{Camilo Fosco}, \bibinfo{person}{Vincent Casser}, \bibinfo{person}{Amish~Kumar Bedi}, \bibinfo{person}{Peter O'Donovan}, \bibinfo{person}{Aaron Hertzmann}, {and} \bibinfo{person}{Zoya Bylinskii}.} \bibinfo{year}{2020}\natexlab{}.
\newblock \showarticletitle{Predicting Visual Importance Across Graphic Design Types}. In \bibinfo{booktitle}{\emph{Proceedings of the 33rd Annual ACM Symposium on User Interface Software and Technology}}.
\newblock


\bibitem[Jiang et~al\mbox{.}(2015)]%
        {jiang2015salicon}
\bibfield{author}{\bibinfo{person}{Ming Jiang}, \bibinfo{person}{Shengsheng Huang}, \bibinfo{person}{Juanyong Duan}, {and} \bibinfo{person}{Qi Zhao}.} \bibinfo{year}{2015}\natexlab{}.
\newblock \showarticletitle{SALICON: Saliency in Context}. In \bibinfo{booktitle}{\emph{IEEE Conference on Computer Vision and Pattern Recognition}}.
\newblock


\bibitem[Jiang et~al\mbox{.}(2023)]%
        {jiang2023ueyes}
\bibfield{author}{\bibinfo{person}{Yue Jiang}, \bibinfo{person}{Luis~A. Leiva}, \bibinfo{person}{Hamed~Rezazadegan Tavakoli}, \bibinfo{person}{Paul~R.B. Houssel}, \bibinfo{person}{Julia Kylm{\"a}l{\"a}}, {and} \bibinfo{person}{Antti Oulasvirta}.} \bibinfo{year}{2023}\natexlab{}.
\newblock \showarticletitle{UEyes: Understanding Visual Saliency across User Interface Types}. In \bibinfo{booktitle}{\emph{Proceedings of the 2023 CHI Conference on Human Factors in Computing Systems}}.
\newblock


\bibitem[Judd et~al\mbox{.}(2012)]%
        {mit300}
\bibfield{author}{\bibinfo{person}{Tilke Judd}, \bibinfo{person}{Fr{\'e}do Durand}, {and} \bibinfo{person}{Antonio Torralba}.} \bibinfo{year}{2012}\natexlab{}.
\newblock \bibinfo{booktitle}{\emph{A Benchmark of Computational Models of Saliency to Predict Human Fixations}}.
\newblock \bibinfo{type}{{T}echnical {R}eport}. \bibinfo{institution}{MIT}.
\newblock


\bibitem[K{\"u}mmerer et~al\mbox{.}(2022)]%
        {deepgaze2021}
\bibfield{author}{\bibinfo{person}{Matthias K{\"u}mmerer}, \bibinfo{person}{Matthias Bethge}, {and} \bibinfo{person}{Thomas S.~A. Wallis}.} \bibinfo{year}{2022}\natexlab{}.
\newblock \showarticletitle{DeepGaze III: Modeling Free-Viewing Human Scanpaths with Deep Learning}.
\newblock \bibinfo{journal}{\emph{Journal of Vision}} \bibinfo{volume}{22}, \bibinfo{number}{5} (\bibinfo{year}{2022}).
\newblock


\end{thebibliography}

\end{document}